\documentstyle[11pt,epsfig,newpasp,twoside]{article}
\def\edcomment#1{\iffalse\marginpar{\raggedright\sl#1\/}\else\relax\fi}
\reversemarginpar
\begin{document}
\title{Jet-Induced Star Formation}
\author{Wil van Breugel, Chris Fragile, Peter Anninos \& Stephen Murray}
\affil{University of California, 
Lawrence Livermore National Laboratory, Livermore,
CA 94550}
\begin{abstract}
Jets from radio galaxies can have dramatic effects on the medium through
which they propagate. We review observational evidence for jet-induced
star formation in low ('FR-I') and high ('FR-II') luminosity radio
galaxies, at low and high redshifts respectively. 
We then discuss numerical simulations
which are aimed to explain a jet-induced starburst ('Minkowski's Object')
in the nearby FR-I type radio galaxy NGC~541.  We conclude that jets can
induce star formation in moderately dense (10 cm$^{-3}$), warm ($10^4$ K)
gas; that this may be more common in the dense environments of forming,
active galaxies; and that this may provide a mechanism for 'positive'
feedback from AGN in the galaxy formation process.
\end{abstract}

\section{Introduction}
Observations suggest that large scale, shock-induced star formation
is an integral part of the galaxy evolution process. It is important
in colliding and merging galaxies such as the Ultra-Luminous Infrared
Galaxies (ULIRGs) and in forming galaxies. It may also occur to a lesser
extent near the central dominant galaxies in cooling flow clusters and
in galaxies moving through dense cluster atmospheres. Perhaps the most
spectacular shock-induced star formation occurs as the result of the
collision of extragalactic radio jets with over-dense ambient material.
Here we review existing observational evidence in support of
jet-induced star formation in both nearby and distant radio galaxies.
We will then discuss numerical simulations performed at the University
of California Lawrence Livermore National Laboratory which are aimed
to explore the conditions under which such star formation may occur,
and apply this to the nearby 'Minkowski's Object' (M.O.).

\section{Environments of radio galaxies}

Radio galaxies do not live in a 'vacuum' but are surrounded by gaseous
halos and/or debris from recent merger events with may have triggered
the radio galaxy activity in the first place. Broadly speaking one can
discriminate between the low-luminosity Fanaroff and Riley FR-I types
(such as 3C~31), which have expanding, turbulent jets without terminal
shocks or hotspots, and the high luminosity FR-II types (such as Cygnus
A) which have very well collimated jets ending at high surface brightness
hotspots and large diffuse lobes (Fanaroff and Riley 1974). FR-I's may be
found in clusters of galaxies where they can interact with cooling clouds
in the Inter Galactic Medium (Ferland et al. 2002).  At high redshift ($z
> 2$), because of sensitivity limits, only the FR-II types are easy to
detect. Since they inhabit young, forming galaxies their jets propagate
through relatively dense and clumpy media.  Observational evidence for
this dense and cool gas in high redshift radio galaxies has been plentiful
and includes the detection of dust, HI, extended line emission, associated
absorption line systems and molecular gas (e.g. De Breuck et al. 2003).

\begin{figure}[h]
\plotone{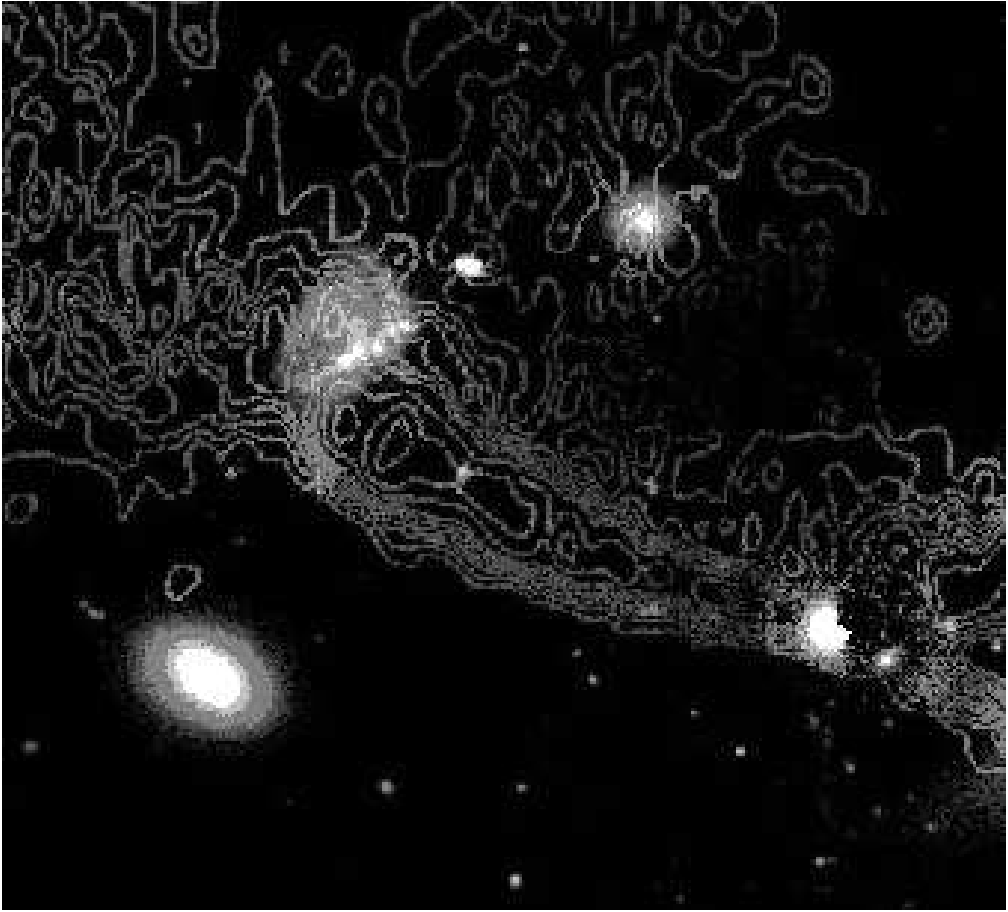}
\caption{Jet-induced starburst in Minkowski's Object, the peculiar object
at the end of the radio jet from NGC~541 (contours, galaxy subtracted;
van Breugel et al. 1985)}
\end{figure}

\subsection{Jet-induced star formation in nearby FR-I type radio galaxies}

One of the first radio galaxies where evidence was found for jet-induced
star formation was the nearest radio galaxy, Centaurus A (Blanco et al
1975).  Further examples have been found as the sensitivity and spatial
resolution of radio and optical telescopes has improved.  In the case
of Centaurus A, recent observations with the Hubble Space Telescope,
when compared to radio images obtained with the Very Large Array,
have confirmed that there are about half a dozen young ($< 15$ Myr)
OB associations near filaments of ionized gas located between the radio
jet and a large HI cloud (Mould et al. 2000).  Another nearby
example is 'Minkowski's Object', a peculiar small starburst system at
the end of a radio jet emanating from the elliptical galaxy NGC~541,
located near the center of the cluster of galaxies Abell~194 (van Breugel
et al. 1985). Star forming regions associated with radio sources have
also been found in cooling flow clusters, with the best example being
in Abell~1795 (McNamara 2002).

\begin{figure}[h]
\plotone{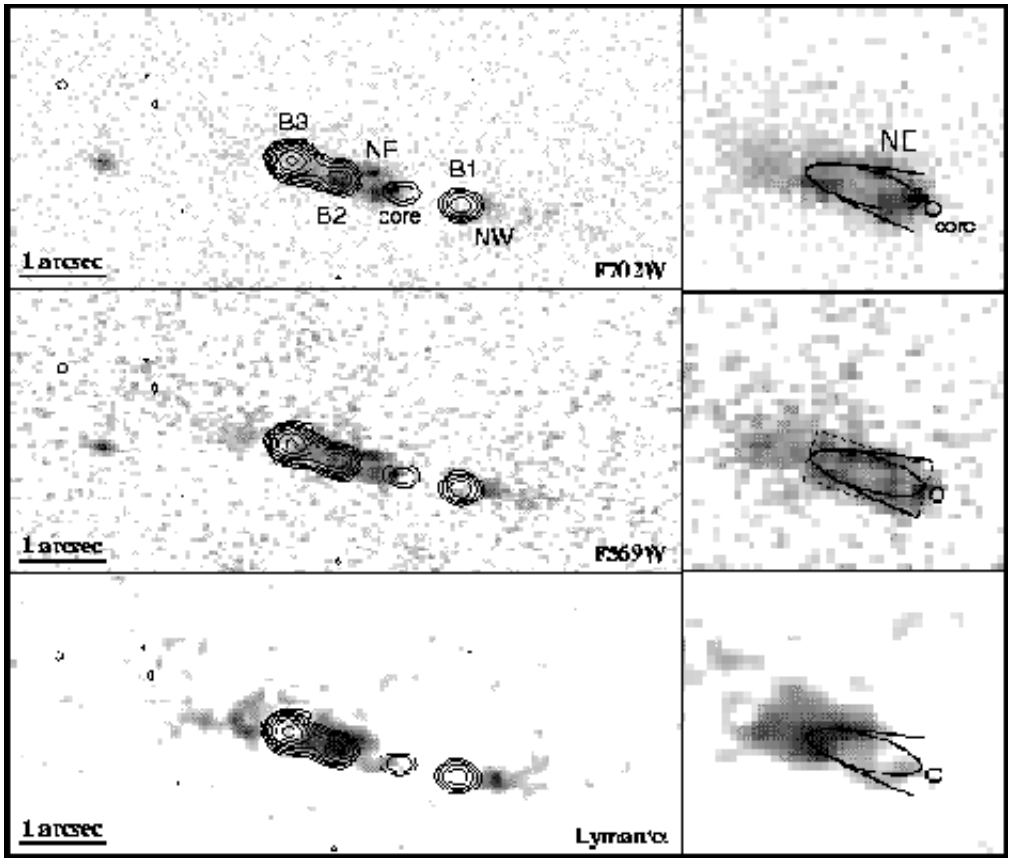}
\caption{Jet-induced star formation in the z = 3.8 FR-II radio galaxy
4C41.17 (Bicknell et al. 2000)}
\end{figure}

\subsection{Jet-induced star formation in distant FR-II type radio galaxies}

Radio galaxies have now been identified up to $z \sim 5.2$. They are the
most massive galaxies at any redshift and, at high redshift, are usually
associated with extended, clumpy systems. Many have giant Ly-a halos,
dust and CO molecular gas. One of the most striking correlations in $z >
0.6$ radio galaxies is that their optical line and continuum emission
is aligned with the radio sources (McCarthy et al. 1987; Chambers et
al. 1987).  The best studied example here is 4C41.17 at $z=3.8$, where
deep spectroscopic observations have shown that the bright, spatially
extended, rest-frame UV continuum emission aligned with the radio axis
of this galaxy is unpolarized and shows P Cygni-like features similar
to those seen in star-forming galaxies (Dey et al. 1977).

Collectively, these observations are best explained by models in which
shocks generated by the radio jet propagate through an inhomogeneous
medium and trigger gravitational collapse in relatively overdense regions
(Begelman et al. 1989; De Young 1989; Rees 1989). A detailed analysis
of the jet-induced star formation in 4C41.17 has been presented by
Bicknell et al. (2000). In that object, Hubble Space Telescope images
showed a bimodal optical continuum structure parallel to the radio jet,
strongly supporting the idea that the star formation was triggered by
sideways shocks.

\begin{figure}[h]
\plotone{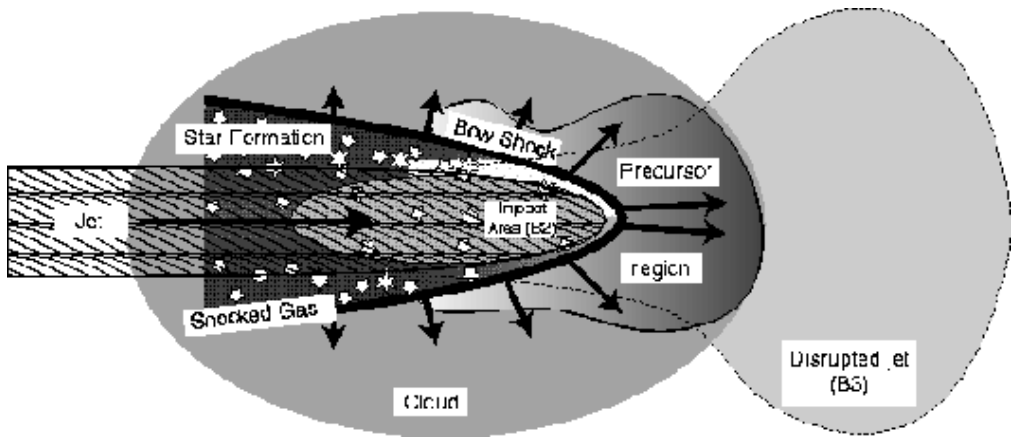}
\caption{Model showing how sideways shocks by expanding radio lobes in an 
FR-II type radio source may induce star formation 
in the dense ambient medium of a forming galaxy (Bicknell et al. 2000)}
\end{figure}

\section{Minkowski's Object}

One of the most spectacular observed jet-induced starbursts is
'Minkowski's Object' associated with the elliptical galaxy NGC~541 in the
cluster of galaxies Abell~194 (van Breugel et al. 1985). Its morphology
is strongly suggestive of a collision between the FR-I type jet from
NGC~541 and a dense cloud: M.O. has the same overall diameter as the
jet, appears wrapped around the end of the jet, has bright emission in
the upstream direction and filamentary structure down-stream where the
jet appears disrupted.  Spectroscopically M.O. looks like an HII region,
resembling starburst galaxies. The H-a luminosity suggests a modest star
formation rate of 0.3 M/yr. VLA observations (van Breugel and van Gorkom,
unpublished) show two detections down-stream from the jet-cloud collision
site, indicating a total HI mass of $\sim 3 \times 10^8$ M$_\odot$.

\section{Numerical simulations}

The basic idea of shock-induced star-formation is that when a strong
shock passes through a clumpy medium it may trigger many smaller-scale
compressive shocks in overdense clumps.  These shocks increase the
density inside the clumps and make it possible for them to radiate more
efficiently.  If the radiative efficiency of the gas has a sufficiently
shallow dependence upon the temperature, then radiative emissions are
able to cool the gas rapidly, in a runaway process, producing even
higher densities as the cooling gas attempts to re-attain pressure
balance with the surrounding medium.  Both the reduction in temperature
and the increase in density act to reduce the Jeans mass, above which
gravitational forces become important. Any clump that was initially
close to this instability limit will be pushed over the edge and forced
into gravitational collapse.  Thus, the passing of a shock through a
clumpy medium may trigger a burst of star formation.  Several processes,
acting on different timescales, govern whether or not cooling and star
formation are able to proceed.

To investigate under which conditions shock-induced star formation can
occur requires numerical simulations. This also provides information
about environments that may be observationally out of reach, such as
in forming galaxies at high redshift, and which could be important
for understanding the role of jets in the feedback of active galactic
nuclei (AGN) on their environment.  Numerical simulations can also
give us insight into the role of shock-induced star formation in other
environments such as supernova shocks and cloud-cloud collisions.

The interaction of a strong shock with non-radiative clouds has been
the subject of many numerical studies (e.g. Klein et al 1994; Poludnenko
et al 2002).  For a non-radiative cloud the passing shock ultimately
destroys the clouds within a few dynamical timescales.  Destruction
results primarily from hydrodynamic instabilities at the interface
between the cloud and the post-shock background gas.

The effects of strong shocks interacting with radiative clouds are very
different and studies of this have only recently begun (Mellema et al. 2002;
Fragile et al. 2003). Instead of re-expanding and quickly diffusing into
the background gas, the compressed cloud instead breaks up into numerous
dense, cold fragments.  These fragments survive for many dynamical
timescales and are presumably the precursors to star formation. Previous
work has focused primarily on FR-II type radio galaxies even though,
because of their proximity, much better observational data can be
obtained for FR-I type jet-induced starbursts such as M.O. Fragile et
al (2003) used the LLNL developed, multi-dimensional, multi-physics,
massively parallel 'COSMOS' numerical simulations package (Anninos et
al. 2003; Anninos and Fragile 2003) to investigate both FR-I and FR-II
type jet-induced star formation systems.

\begin{figure}[h]
\plotone{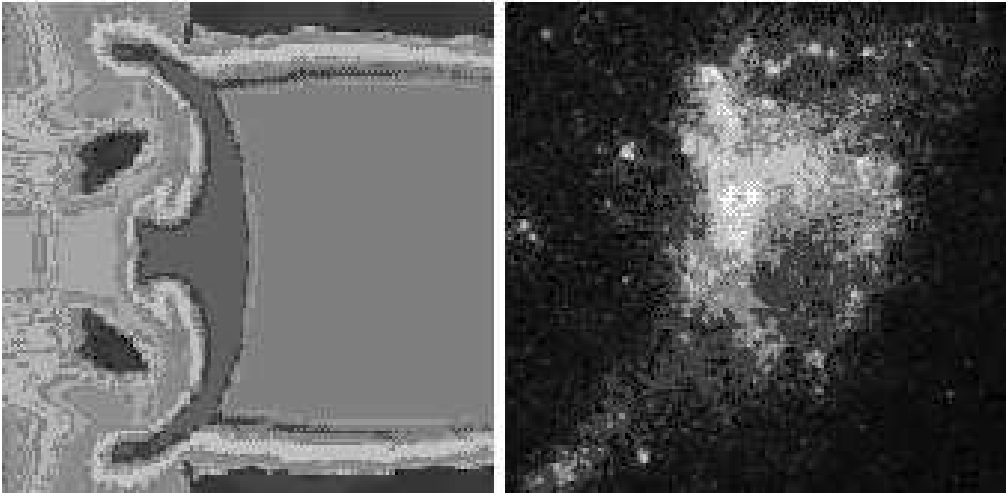}
\caption{Comparison of an intermediate density distribution plot from the
numerical simulation with a similarly scaled observation of Minkowski's
Object (rotated for easy comparison).  There are clear similarities
between the distribution of the post-shock gas within the simulated
cloud and the regions of active star formation within Minkowski's Object.
}
\end{figure}

\begin{figure}[ht]
\plotone{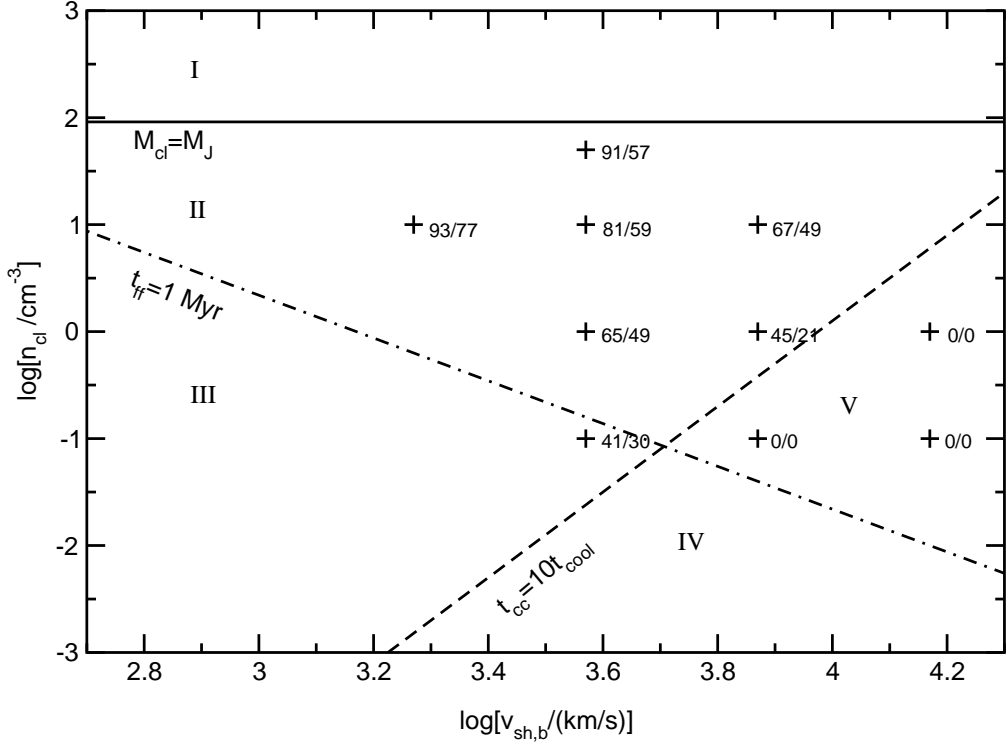}
\caption{General cloud number density ($n_cl$)-shock velocity ($v_{sh,b}$)
parameter space considered.  The solid line demarks the
density above which the cloud is initially graviationally unstable
(region I).  The dashed line divides the cooling dominated regions
(II \& III) on the left from the non-cooling regions (IV \& V) on the
right.  The dot-dashed line is an estimate of the star-formation
cut-off.  
The parameter pairs explored
with the equilibrium cooling curve model are indicated
with crosses.  The numbers give the percent of the initial cloud mass that 
ends up below $T=1000$ and 100 K, respectively.
}
\end{figure}

\subsection{Application to Minkowski's Object}

To simulate the jet induced star formation in M.O. Fragile et al (2003)
assumed that NGC~541 is surrounded by a multi-phase medium resembling
cluster atmospheres (e.g. Ferland et al. 2002).  Specifically, it was
assumed that the FR-I jet interacts with a moderate density,  
hot `mother'-cloud with a semimajor axis of 10 kpc, a semiminor axis of
5 kpc, a density $n_{mcl} = 0.1$ cm$^{-3}$, and temperature $T=10^6$ K.
This corresponds to an initial total cloud mass of $\approx 10^9 M_\odot$.
Within this mother-cloud denser, warm clouds were assumed to be embedded
with typical sizes of 100 pc, $n_{cl} = 10$ cm$^{-3}$, and temperature
$T=10^4$ K.

The detailed radio and X-ray study of the proto-typical FRI-type radio
galaxy 3C31 by Laing and Bridle (2002) was used to estimate plausible jet
parameters near M.O., at $\sim 15$ kpc from the NGC~541 AGN.  We assumed
that the long axis of the mother-cloud is aligned with a jet of high
velocity ($0.3 c$), low density ($10^{-4}$ cm$^{-3}$) gas flowing onto
the grid.  The diameter of the jet nozzle is equal to half the diameter of
the cloud along the semiminor axis. 

Our numerical simulations consisted of two parts. To investigate the
overall structure of M.O. we followed the evolution of the collision of
the jet with the low density mother-cloud. To determine whether star
formation can be triggered we followed the evolution of shocks interacting
with one or more of the dense, warm clouds embedded in the mother-cloud.

The collision with the mother-cloud triggers a nearly
planar shock down the long axis of the cloud.  As the bow shock from the
jet wraps around, it also triggers shocks along the sides of
the cloud.  A similar shock structure may explain the filamentary nature
of the star-forming region in M.O. (Fig. 4).
To determine whether the clouds embedded within the mother-cloud would
indeed collapse and form stars we ran a number of simulations to explore
the density / velocity parameter space. These results are summarized
in Figure 5.

\section{Conclusions}

From our numerical simulations, when applied to M.O., a number of
important conclusions can be drawn. First, its peculiar morphology
- bright star forming region orthogonal to the jet, and fainter
filamentary features downstream from there - can be easily reproduced
by our simulations.  Second, the modest amount
star formation required - $0.3 M_\odot$ yr$^{-1}$ for the entire object,
is also easily achieved for the plausible parameter space explored by
our simulations.  Third, and most interestingly, we conclude that the
star formation in M.O. could be induced by a moderate
velocity jet ($9 \times 10^4$ km s$^{-1}$) interacting with a collection
of slightly overdense ($\sim 10$ cm$^{-3}$), warm ($10^4$ K) clouds,
i.e. it is NOT necessary to assume that this was an accidental collision
between a jet and a preexisting gas-rich galaxy. This also suggests that
the neutral hydrogen associated with M.O. ($3 \times 10^8
M_\odot$; W. van Breugel \& J. van Gorkom 2003, private communication)
may have cooled from the warm gas phase as a result of the radiative
cooling triggered by the radio jet.  An update on the observations of
M.O. will be presented in a forthcoming paper (S. D. Croft
et al., in preparation).

Our models can be used to infer the importance of shock-induced star
formation in other regimes by simply adjusting the criteria used to
develop Figure 5. They have confirmed the reality of the division between
regions II and V (cooling and non-cooling) derived using approximate
analytical methods, while the other region boundaries are set by physical
limits for the clouds. Our key conclusion is that shocks associated with
jets may indeed trigger the collapse of clouds to form stars. Whether
this occurs at the impact area, or along the sides of expanding lobes
depends on jet power and the ambient gas density distribution. In
forming galaxies, where both dense gas and AGN activity are likely,
jet-induced star formation might help trigger even more star formation,
at an earlier stage, then otherwise might have occured.

In future work, we shall expand
our numerical simulations package to include Adaptive Mesh Refinement
and magnetic fields and will investigate regimes appropriate for other
possible shock-cloud interaction scenarios.

\acknowledgments{This work was performed under the auspices of the
U.S. Department of Energy by University of California, Lawrence
Livermore National Laboratory under Contract W-7405-Eng-48.  W.v.B.\
also acknowledges NASA grants GO~9779 and GO3-4150X in support of
high-redshift radio galaxy research with HST and Chandra.
}

\end{document}